\documentclass[aps,manuscript,showpacs,showkeys,superscriptaddress]{revtex4-1}
\usepackage{graphicx}
\usepackage{epsfig}  
\usepackage{epsf}    
\usepackage{dcolumn}
\usepackage{bm}
\usepackage{dcolumn}
\usepackage{textcomp}
\usepackage[tbtags]{amsmath}
\usepackage{amsfonts}
\usepackage{float}
\usepackage{subfig}
\usepackage[]{hyperref}
  \hypersetup{
  unicode=false,          
  pdftoolbar=true,        
  pdfmenubar=true,        
  pdffitwindow=true,     
  pdfstartview={FitH},    
  pdfsubject={Getting the best out of T2K and NOvA},   
  pdfnewwindow=true,      
  pdfcreator={RevTeX},
  colorlinks=true,       
  linkcolor=red,          
  citecolor=blue,        
  urlcolor=blue,           
  }
\usepackage{hypcap}

\newcommand{\beq}{\begin{equation}}
\newcommand{\eeq}{\end{equation}}
\newcommand{\beqa}{\begin{eqnarray}}
\newcommand{\eeqa}{\end{eqnarray}}

\newcommand{\tx}{{\theta_{12}}}
\newcommand{\ty}{{\theta_{13}}}
\newcommand{\tz}{{\theta_{23}}}

\newcommand{\atil}{\hat{A}}
\newcommand{\dtil}{\hat{\Delta}}

\newcommand{\dcp}{\delta_{\mathrm{CP}}}
\newcommand{\nova}{NO$\nu$A~}
\newcommand{\pmue}{P(\nu_\mu \rightarrow \nu_e)}
\newcommand{\pme}{P_{\mu e}}
\newcommand{\pmuebar}{P(\bar{\nu}_{\mu} \rightarrow \bar{\nu}_e)}
\newcommand{\pmebar}{P_{\bar{\mu} \bar{e}}}

\begin{document}


\title{Tensions between the appearance data of T2K and \nova}
\author{Mohammad Nizam}
\email[Email Address: ]{mohammad.nizam@tifr.res.in}
\affiliation{Tata Institute of Fundamental Research,
Mumbai 400005, India}
\affiliation{Homi Bhabha National Institute, Anushakti Nagar, 
Mumbai 400094, India}
\author{Suman Bharti}
\email[Email Address: ]{sbharti@phy.iitb.ac.in}
\affiliation{Department of Physics, Indian Institute of Technology Bombay,
Mumbai 400076, India}
\author{Suprabh Prakash}
\email[Email Address: ]{sprakash@ifi.unicamp.br}
\affiliation{Instituto de F\'isica Gleb Wataghin - UNICAMP, 13083-859, Campinas, SP, Brazil}
\author{Ushak Rahaman}
\email[Email Address: ]{ushak@phy.iitb.ac.in}
\affiliation{Department of Physics, Indian Institute of Technology Bombay,
Mumbai 400076, India}
\affiliation{Centre for Astro-Particle Physics (CAPP) and Department of Physics, 
University of Johannesburg, PO Box 524, Auckland Park 2006, South Africa}
\author{S. Uma Sankar}
\email[Email Address: ]{uma@phy.iitb.ac.in}
\affiliation{Department of Physics, Indian Institute of Technology Bombay,
Mumbai 400076, India}
\date{\today}
\begin{abstract}
The long baseline neutrino experiments, T2K and NO$\nu$A, have taken significant amount of data in each of the 
four 
channels: (a) $\nu_\mu$ disappearance, (b) $\bar\nu_\mu$ disappearance (c) $\nu_e$ appearance and (d) $\bar\nu_e$ 
appearance. There is a 
mild tension between the disappearance and the appearance data sets of T2K. A more serious tension exists between the 
$\nu_e$ appearance data of T2K and the $\nu_e / \bar\nu_e$ appearance data of NO$\nu$A. 
This tension is significant enough that T2K rules out the best-fit point of 
\nova at $95\%$ confidence level whereas \nova rules out T2K best-fit point at $90\%$ confidence level. We explain 
the reason why these tensions arise. We also do a combined fit of T2K and 
\nova data and comment on the results of this fit.
\end{abstract}
\maketitle

\section{Introduction}
Recently T2K \cite{Abe:2018wpn} and \nova \cite{Jun-2018} collaborations have 
published their data on both 
neutrino and anti-neutrino runs. The best-fit values of various neutrino 
oscillations parameters determined by each of these experiments are listed 
in table~\ref{best-fit-data-table} below. In the table, NH refers to normal 
hierarchy ($\Delta_{32}> 0$) and IH refers to inverted hierarchy ($\Delta_{32}< 
0$). 

\begin{table}[htbp]
\begin{center}
 \begin{tabular}{|l|l|l|}
\hline \hline
    Parameter & \nova & T2K  \\ 
\hline \hline
   $\Delta_{32}/10^{-3} {\rm\ eV^2\ (NH)}$   & 2.51   & 2.463  \\ 
\hline
   $\Delta_{32}/10^{-3} {\rm\ eV^2\ (IH)}$   & -2.56   & -2.506  \\ 
\hline
   $\sin^2\theta_{23} {\rm\ (NH)}$          & 0.58   & 0.526 \\
\hline
   $\sin^2\theta_{23} {\rm\ (IH)}$          & 0.58   & 0.530 \\
\hline
   $\dcp {\rm\ (NH)}$  & $30.6^\circ$   & $-107.1^\circ$ \\
\hline
   $\dcp {\rm\ (IH)}$  & $-95.4^\circ$   & $-81.9^\circ$ \\
\hline \hline  
  \end{tabular}
 \caption{\footnotesize{Best-fit points of T2K \cite{Abe:2018wpn} and 
\nova \cite{Jun-2018} data. The fit is based on the following numbers of 
protons on target (POT). For \nova POT are $8.85\times 10^{20}$ 
($6.9\times 10^{20}$) in neutrino (anti-neutrino) modes. For T2K they are 
$14.7\times 
10^{20}$ ($7.6\times 10^{20}$) in neutrino (anti-neutrino) modes.}}
\label{best-fit-data-table}
\end{center}
\end{table}

The best-fit values of $\Delta_{32}$, both for NH and for IH, 
for the 
two experiments are very close. These two experiments are sensitive to all the three unknown parameters of neutrino 
oscillations: the neutrino 
mass hierarchy, the octant of $\theta_{23}$ and the CP violating phase $\dcp$. Both experiments 
prefer NH over IH and the higher octant for $\theta_{23}$. However, T2K prefers $\sin^2\theta_{23}$ close to the 
maximal value of $0.5$ whereas \nova prefers a higher value which is closer to $0.6$. Regarding $\dcp$, T2K demands 
that it should be in the lower half plane (LHP) with a preference for a value close to $-90^\circ$. The 
best-fit value of $\dcp$ for \nova is in the upper half plane (UHP) and it disfavours values close to $-90^\circ$.  
The 
results of these experiments are 
usually given in the form of contours of allowed regions in $\dcp-\sin^2\theta_{23}$ plane, for NH and for IH. 
Overall, 
T2K rules out the best-fit point of \nova in this plane at $95\%$ confidence level and \nova 
rules out the T2K best-fit point at $90\%$ confidence level. This disagreement is the result of the tension between 
the appearance data of these two experiments. Though both the experiments prefer NH, the possibility of IH is not 
ruled out.
The IH best-fit point of T2K is allowed only at $2~\sigma$ while the 
corresponding point for \nova is allowed at $1~\sigma$. It is interesting to 
note that the IH best-fit points of the two experiments are reasonably 
close to each other.

Long baseline accelerator neutrino experiments take data in the following four channels:
\begin{itemize}
 \item \underline{$\nu_\mu$ disappearance} The number of events in this channel are the largest because the survival 
probability is moderately large over a wide energy range and also because the neutrino flux and the cross section are 
larger. Hence this channel has the highest statistical weight.
 \item \underline{$\bar{\nu}_\mu$ disappearance} This channel has the second highest statistical weight. The survival 
probability is moderately large but the neutrino flux and the cross section are smaller.
 \item \underline{$\nu_e$ appearance} The statistical weight of this channel is not very high because the oscillation 
probability is rather small ($\leq 0.05$).
 \item \underline{$\bar{\nu}_e$ appearance} This channel has the least statistical weight because the oscillation 
probability is low and the anti-neutrino flux and the cross section are smaller. 
\end{itemize}
The tension between the 
data of the two experiments may well be the result of low statistics and may disappear with more data. However, it is 
possible to make some predictions regarding the trend along which the tension is likely to be resolved by doing a 
combined fit.

In this article we study the neutrino and anti-neutrino 
appearance data of these two experiments using the expressions for $\pmue$ and 
$\pmuebar$ in three flavour oscillations including 
matter effect. The disappearance data is related to the survival probabilities $P(\nu_\mu\rightarrow\nu_\mu)$ and 
$P(\bar\nu_\mu\rightarrow\bar\nu_\mu)$. For the energies and the baselines of T2K and NO$\nu$A, the survival 
probabilities 
can be approximated by a two flavour expression with an effective mass-squared difference 
$\Delta_{\mu\mu}$~\cite{Nunokawa:2005nx} and $\sin 2\tz$. We find that there are mild tensions between the neutrino 
and 
anti-neutrino disappearance data of \nova\cite{Jun-2018} and between the neutrino 
appearance and disappearance data of T2K. These tensions are likely to be resolved with more data. There is, 
however, 
a serious tension between the appearance data of the two experiments, which is leading to each experiment ruling out 
the best-fit point of the other experiment. We explain the cause of this tension. We 
also do a combined fit of the data from both the experiments and compare the best-fit point to the current 
best-fit points of the two experiments.

\section{$\pme$ and $\pmebar$ for T2K and NO$\nu$A }
For both T2K and \nova experiments, the expression for $\pmue$ is 
\cite{Cervera:2000kp,Freund:2001pn}
\begin{eqnarray}
\pmue & = & \pme =
\sin^2 2 \ty \sin^2 \tz\frac{\sin^2\dtil(1-\atil)}{(1-\atil)^2}+ \nonumber\\ 
& & \alpha \cos \ty \sin2\tx \sin 2\ty \sin 2\tz \cos(\dtil+\dcp)
\frac{\sin\dtil \atil}{\atil} \frac{\sin \dtil(1-\atil)}{1-\atil},
\label{pmue-exp}
\end{eqnarray}
where $\hat{\Delta} = 1.27 \Delta_{31}L/E$, $\hat{A} = A/\Delta_{31}$ and 
$\alpha = \Delta_{21}/\Delta_{31}$.
The Wolfenstein matter term $A$ is \cite{msw1}
\begin{equation}
A \ ({\rm in \ eV^2}) = 0.76 \times 10^{-4} \rho \ ({\rm in \ gm/cc}) \ E \ ({\rm in \ GeV}),
\end{equation}
where $E$ is the energy of the neutrino and $\rho$ is the density of the matter.
For anti-neutrinos, 
$P(\bar{\nu}_\mu \to \bar{\nu}_e) = \pmebar$ is given by a similar expression
with $\dcp \to - \dcp$ and $A \to -A$. Since $\alpha\approx 0.03$, the term 
proportional to $\alpha^2$ in $\pme$ is neglected.

The best-fit points of T2K and NO$\nu$A, for both NH and IH, can be understood 
by considering the 
changes induced in $\pme$ and $\pmebar$ by the change in each of the unknowns 
relative to a 
common reference set of parameter values. We take this reference set to be  
vacuum oscillations with $\tz=45^\circ$ and $\dcp=0$.
\begin{itemize}
\item Inclusion of matter effect increases 
$\pme$ for NH and decreases it for IH. The effect is opposite for $\pmebar$.
\item  Both $\pme$ and $\pmebar$ increase if $\tz$ is in the higher octant (HO) 
and decrease if it is in the lower octant (LO).
\item If $\dcp$ is in the lower half plane (LHP) $\pme$ increases whereas it 
decreases for $\dcp$ in the upper half plane (UHP). Here again the effect is 
opposite for $\pmebar$. 
\end{itemize}

If the hierarchy is NH, the octant is HO and $\dcp$ is in LHP, all the three unknowns boost $\pme$ and we expect 
a large excess of $\nu_e$ appearance events. This combination, however, leads to a moderate suppression of $\pmebar$ 
because it is reduced due to hierarchy and $\dcp$ and increased due to octant. A moderate increase in $\pme$, and 
hence in $\nu_e$ appearance events, occurs when two unknowns boost it and the third suppresses it. Three different 
combinations can cause this possibility. They are: 
\begin{itemize}
 \item (A) Hierarchy is NH, octant is HO and $\dcp$ is in UHP. For this combination, the effects of hierarchy and 
$\dcp$ are opposite for both $\pme$ and $\pmebar$. Both these probabilities receive a modest boost due to HO.
\item (B) Hierarchy is NH, octant is LO and $\dcp$ is in LHP. Here hierarchy and $\dcp$ both boost $\pme$ and LO 
lowers it. All three parameters lower $\pmebar$ leading to the lowest expected number of $\bar\nu_e$ appearance 
events.
\item (C) Hierarchy is IH, octant is HO and $\dcp$ is in LHP. Here again, the effects of hierarchy and $\dcp$ are 
opposite for both $\pme$ and $\pmebar$ and both these probabilities receive a modest boost due to HO. 
\end{itemize}

For the T2K experiment, the peak flux occurs for $E_\nu\approx 0.6$ GeV. Hence the matter effects are small and lead 
to a change of about $8\%$ in $\pme$ and $\pmebar$. Maximal values of $\dcp$ can change the probability by about 
$20\%$. The change induced by the octant, of course depends on the value of $\sin^2\tz$. For the \nova experiment, 
the flux peaks at $E_\nu\approx 2$ GeV. The corresponding matter effects change probability by about $20\%$, which 
is also the change induced by maximal CP violation. The first neutrino data of \nova had the following 
features: (a) $\nu_\mu$ disappearance preferred non-maximal $\tz$ and (b) $\nu_e$ appearance showed a modest increase 
relative to the expectation from the reference point~\cite{Adamson:2017gxd}. The non-maximal $\tz$ values also induced 
a $20\%$ change in $\pme$ and $\pmebar$. Thus, each of the three unknowns induced change of similar magnitude which 
lead to three degenerate solutions of the forms (A), (B) and (C) listed above~\cite{Bharti:2018eyj}.

\section{Current accelerator neutrino data}
\subsection{T2K}
T2K experiment observed maximal disappearance in both $\nu_\mu$ and $\bar\nu_\mu$ channels. This implies that 
$\sin^2\theta_{23}$ is close to $0.5$. 
We did separate analyses of the disappearance data and the appearance data of T2K. In these analyses,
the theoretical expectations are 
calculated using the software GLoBES~\cite{Huber:2004ka,Huber:2007ji}. We matched the 
GLoBES predictions for the expected bin-wise event numbers with 
those given by the 
Monte-Carlo simulations of the experiments, quoted in 
refs.~\cite{Abe:2018wpn} and~\cite{Jun-2018}, for the same input parameters. 
In calculating the theoretical expectations, the values of $\Delta_{21} = 
7.50\times 10^{-5}$ eV$^2$ and $\sin^2\tx = 0.307$ are held fixed. The other 
mixing angles are varied in the following ranges: $\sin^22\ty = (0.084\pm3\times 0.003)$ and $\sin^2\tz = 
(0.25,0.75)$. The CP-violating phase is varied over its full range $\dcp = (-180^\circ,180^\circ)$. The effective 
mass-squared 
difference, $\Delta_{\mu\mu}$~\cite{Nunokawa:2005nx}, is varied in the range 
$(2.32\pm 3\times 0.11)\times 10^{-3}$ eV$^2$.
The value of $\Delta_{31}$ is determined from the equation~\cite{Nunokawa:2005nx} 
\begin{equation}
 \Delta_{31}=\Delta_{\mu\mu}+\Delta_{21}(\cos^2\tx-\cos\delta \sin\ty \sin2\tx \tan\tz).
\end{equation}
For NH, $\Delta_{\mu\mu}$ is positive and for IH it is negative. 
 
The results of the disappearance analysis give the best-fit value of $\sin^2\tz$ 
to be $0.51$.
The data from both neutrino and 
anti-neutrino channels are included in this analysis.  We also find that this data constrains $\sin^2\tz$ to be in the 
range 
$(0.43, 0.6)$ at $3~\sigma$. The disappearance data has no sensitivity to $\dcp$. Therefore, the constraints on 
$\sin^2\tz$ are valid for all values of $\dcp$.

\begin{figure}[t]
\centering
\includegraphics[width=1.0\textwidth]{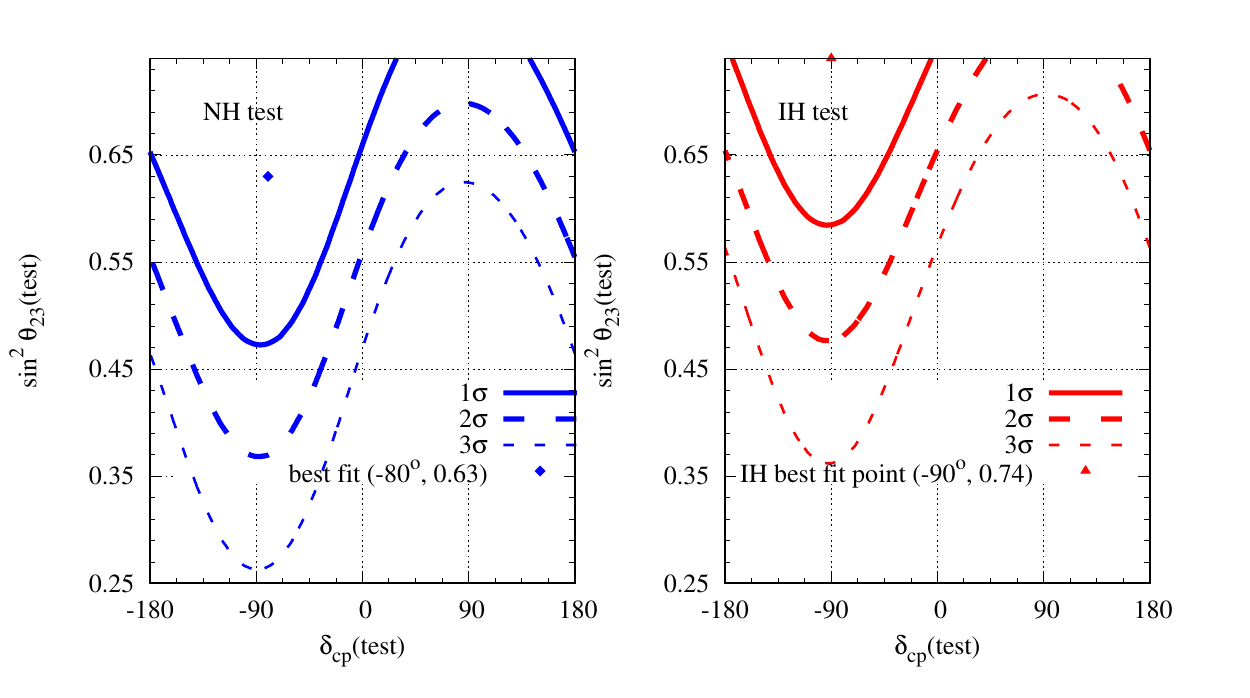}
\caption{\footnotesize{Expected allowed regions in 
$\sin^2\theta_{23}-\dcp$ plane from the current appearance data of 
T2K in the neutrino channel. In the left panel, the 
hierarchy is assumed to be NH and in the right panel, the hierarchy is assumed  
to be IH. The best-fit point has minimum $\chi^2=29$ for both NH and IH for 
$24$ energy bins.  
}}
\label{T2K-nue-app}
\end{figure}

From table-2 of reference~\cite{Abe:2018wpn} we can 
estimate that the $\nu_e$ 
appearance events for the reference point is about $60$, for the given neutrino 
run of T2K. Inclusion of matter effects changes this number by $4$ and inclusion of maximal CP violating effects 
changes it by about $11$. Thus, for T2K, the change induced by CP violation is much larger than the change induced 
by the matter effects. The combined effect of   
matter effects (assuming NH) and CP violation effects (assuming $\dcp=-90^\circ$) increases the estimated 
number to $80$~\cite{Abe:2018wpn}. Any further increase must necessarily require a value of $\sin^2\tz$ larger than 
$0.5$. T2K observes $89$ $\nu_e$ appearance events. Hence, the $\nu_e$ appearance data of T2K pulls $\sin^2\tz$ to 
larger 
values. An analysis of the $\nu_e$ appearance data is shown in fig~\ref{T2K-nue-app}. It gives the best-fit value of 
$\sin^2\tz$ as $0.63$, 
though $0.5$ is  
allowed within the allowed $1~\sigma$ range. Thus, we see that there is a mild tension between T2K disappearance data 
and its $\nu_e$ appearance data. This tension may be due to the limited statistics and may go away with more data. 
However, the final value of $\sin^2\tz$ is likely to be determined by the disappearance data because of its
larger statistical weight. The combined analysis done by the T2K collaboration gives the best-fit value 
$\sin^2\tz=0.53$~\cite{Abe:2018wpn}. In doing the analysis of the appearance data we have not included the data from 
the anti-neutrino channel. The number of events observed in this channel are too small for any meaningful analysis. 

The observed large excess of $\nu_e$ events requires  
that $\dcp$ must be in the neighbourhood of $-90^\circ$. For values of $\dcp$ in the UHP, the expected number of 
$\nu_e$ appearance events is smaller than the estimated 
number for the reference point. Since the observed number is much higher, $\dcp$ in UHP is 
strongly disfavoured. 
This data also 
disfavours IH because the corresponding matter effects lead to a lower prediction for the number of events. IH with 
$\dcp=-90^\circ$ is barely allowed at $2~\sigma$~\cite{Abe:2018wpn}. 

T2K has also observed $7$ $\bar\nu_e$ appearance events. For $\dcp=0$ and NH, they expect to observe 
$9$ events. This 
number is expected to go down to $8$ if $\dcp=-90^\circ$. Therefore, the number of $\nu_e$ and $\bar\nu_e$ appearance 
events are consistent with each other. Due to the large statistical error in the $\bar\nu_e$ appearance events, it is 
not possible to make any strong comment on the value of $\dcp$ preferred by this data.


\subsection{NO$\nu$A}

\begin{figure}[t]
\centering
\includegraphics[width=1.0\textwidth]{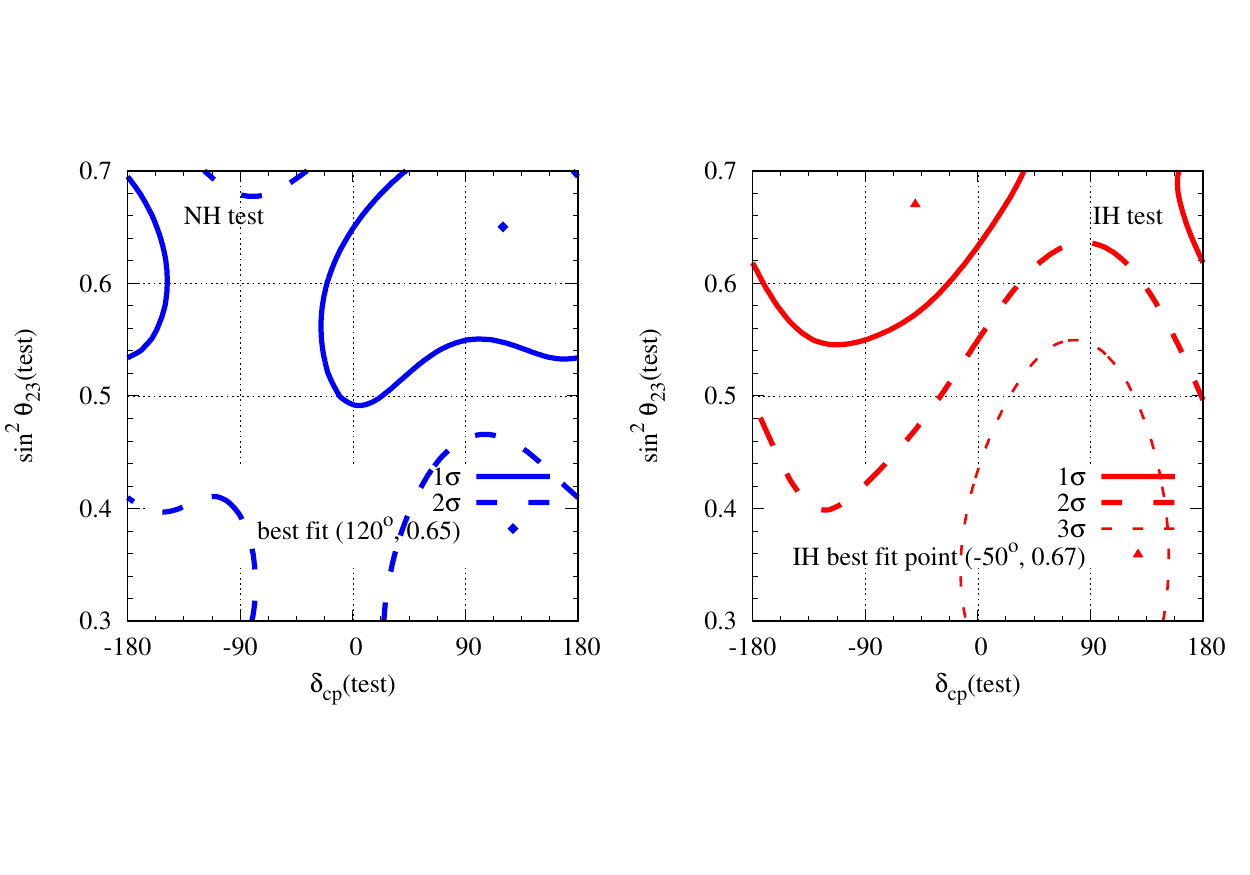}
\caption{\footnotesize{Expected allowed regions in 
$\sin^2\theta_{23}-\dcp$ plane from the appearance data 
of \nova in both neutrino and anti-neutrino channels, as given in~\cite{Jun-2018}. In the left panel, the 
hierarchy is assumed to be NH and in the right panel, the hierarchy is assumed  
to be IH. The best-fit point occurs for NH with a minimum $\chi^2=6$ for $12$ data 
points. The best-fit point of IH has $\Delta\chi^2=0.3$.   
}}
\label{nova-app}
\end{figure}

The recent neutrino disappearance data of \nova is consistent with maximal mixing ($\sin^2\tz = 0.5$) whereas the 
anti-neutrino disappearance data prefers a non-maximal value~\cite{Jun-2018}. Here again there is a 
mild tension between the different data sets of the same 
experiment. This tension also is not statistically significant because of the limited statistics of the anti-neutrino 
data.
Regarding the appearance events, we expect 
$39$ $\nu_e$ events and $15.5$ $\bar\nu_e$ events for the reference point~\cite{Bharti:2018eyj}. 
NO$\nu$A experiment observed $58$ and $18$ events respectively in these channels. That is, there is a moderate excess 
in both these channels. As mentioned above, each of the three unknowns induce  
a change of about $20\%$ in the appearance events of NO$\nu$A. A moderate excess in both $\nu_e$ and $\bar\nu_e$ 
appearance events is possible only if the changes induced by the hierarchy and $\dcp$ cancel each other and the 
increase in both channels occurs because $\tz$ is in HO. That is, the combination of the unknowns must have either 
form (A) or form (C) listed in section-2. We performed an analysis of \nova appearance data in a 
manner similar to the analysis we did for T2K data. The results are shown in fig.~\ref{nova-app}. There are two 
nearly degenerate best-fit solutions, with the unknown parameter values (NH, $\sin^2\tz=0.65$, $\dcp=120^\circ$) 
and (IH, $\sin^2\tz=0.67$, $\dcp=-50^\circ$) respectively. The first solution is in the form (A) and the second in 
the form (C).

The analysis done by the \nova collaboration, of both their disappearance and appearance data, also finds two nearly 
degenerate 
best-fit points~\cite{Jun-2018}: 
(NH, $\sin^2\tz=0.58$, $\dcp=30.6^\circ$) and (IH, $\sin^2\tz=0.58$, $\dcp=-90^\circ$). Here again, we find two 
solutions, one in form (A) and the other in form (C). Since the disappearance data also is included in the \nova 
analysis, a smaller value of $\sin^2\tz$ is obtained compared to the values shown in fig.~\ref{nova-app}. The 
wide difference in the values of $\dcp$ preferred by T2K 
and by \nova shows the tension
between the data of \nova and T2K. T2K prefers a large boost of $\pme$ by all the three unknowns whereas 
\nova prefers a moderate boost of both $\pme$ and $\pmebar$ due to the combinations mentioned above. This tension is 
also visible in the fact that T2K rules out the best-fit 
point of \nova at $95\%$ C.L. whereas \nova rules out the best-fit point of T2K at $90\%$ C.L. The appearance 
events of NO$\nu$A, especially in the $\bar\nu_e$ channel, are limited. With more statistics it is possible that the 
present tension may go away. But the likely resolution of this tension will have $\dcp$ close to $-90^\circ$ because 
of the very large excess of $\nu_e$ appearance events observed by T2K.

\section{Combined fit to T2K and \nova data}

\begin{figure}[t]
\centering
\includegraphics[width=1.0\textwidth]{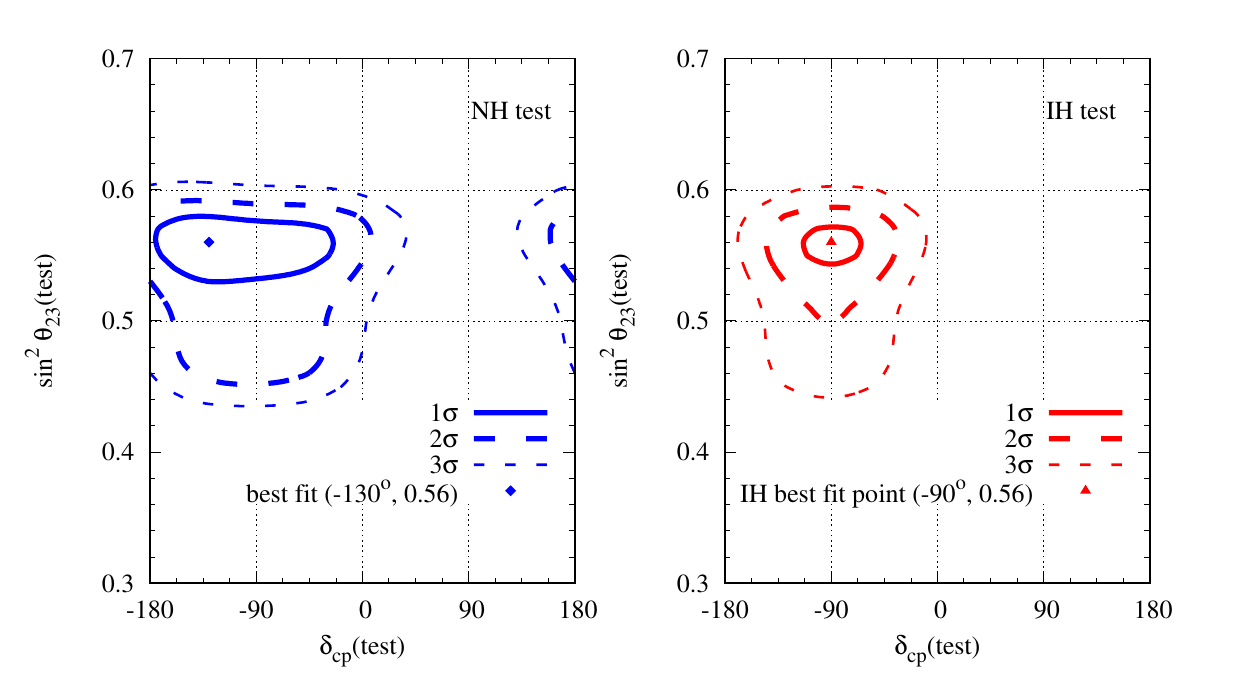}
\caption{\footnotesize{Expected allowed regions in 
$\dcp-\sin^2\theta_{23}$ plane from the combined fit of the neutrino and 
anti-neutrino data of T2K and NO$\nu$A, as of July 2018. In the left panel, the 
hierarchy is assumed to be NH and in the right panel, the hierarchy is assumed  
to be IH. The $\chi^2$ for NH best-fit point is $219$ and that for IH best-fit point is $220.5$, for $182$ 
data points.   
}}
\label{newsys10}
\end{figure}

In this section we present our results of combined fit of the disappearance 
and the appearance data 
of T2K and \nova in both neutrino and anti-neutrino channels. The data of T2K 
is taken from ref.~\cite{Abe:2018wpn} and that of 
\nova from ref.~\cite{Jun-2018}. 
The theoretical expectations for the two experiments are 
calculated using the software GLoBES~\cite{Huber:2004ka,Huber:2007ji}, using the procedure that was described in 
section-3.1 for T2K analyses. There are a total of $182$ energy bins, half in neutrino channel and half in 
anti-neutrino channel. In each case, there are $42$ in disappearance data and $24$ in appearance data 
for T2K and $19$ in disappearance data and $6$ in appearance data for NO$\nu$A.  
In computing the $\chi^2$ between the data and the theoretical expectations, 
prior is added for $\sin^22\ty$ and $\Delta_{\mu\mu}$. We have also included a $10\%$ overall systematic error for 
each channel of 
both the 
experiments.
The results of our fit are shown in figure~\ref{newsys10}.  The best-fit point for NH is ($\sin^2\tz=0.56$, 
$\dcp=-130^\circ$)  
 with a $\chi^2$ of $219$ and the best-fit point for IH is ($\sin^2\tz=0.56$, 
$\dcp=-90^\circ$) with a $\chi^2$ of $220.5$.  

The best-fit point for NH seems to be a compromise between the best-fit points of T2K and NO$\nu$A, especially with 
regard to choice of $\sin^2\tz=0.56$. On the other hand, the choice of $\dcp = -130^\circ$ as its best-fit 
value is enforced by the large excess of $\nu_e$ events observed by T2K. For NH, values of $\dcp$ in UHP 
are essentially ruled out at $2~\sigma$.  
For IH, the best-fit point in our fit is close to the IH best-fit points of T2K 
and NO$\nu$A. This is 
not surprising because those two points are close to each 
other. For IH, the whole region of $\dcp$ in upper half plane is ruled 
out at $3~\sigma$ because it is disfavoured by both T2K and NO$\nu$A.

\begin{figure}[t]
\centering
\includegraphics[width=1.0\textwidth]{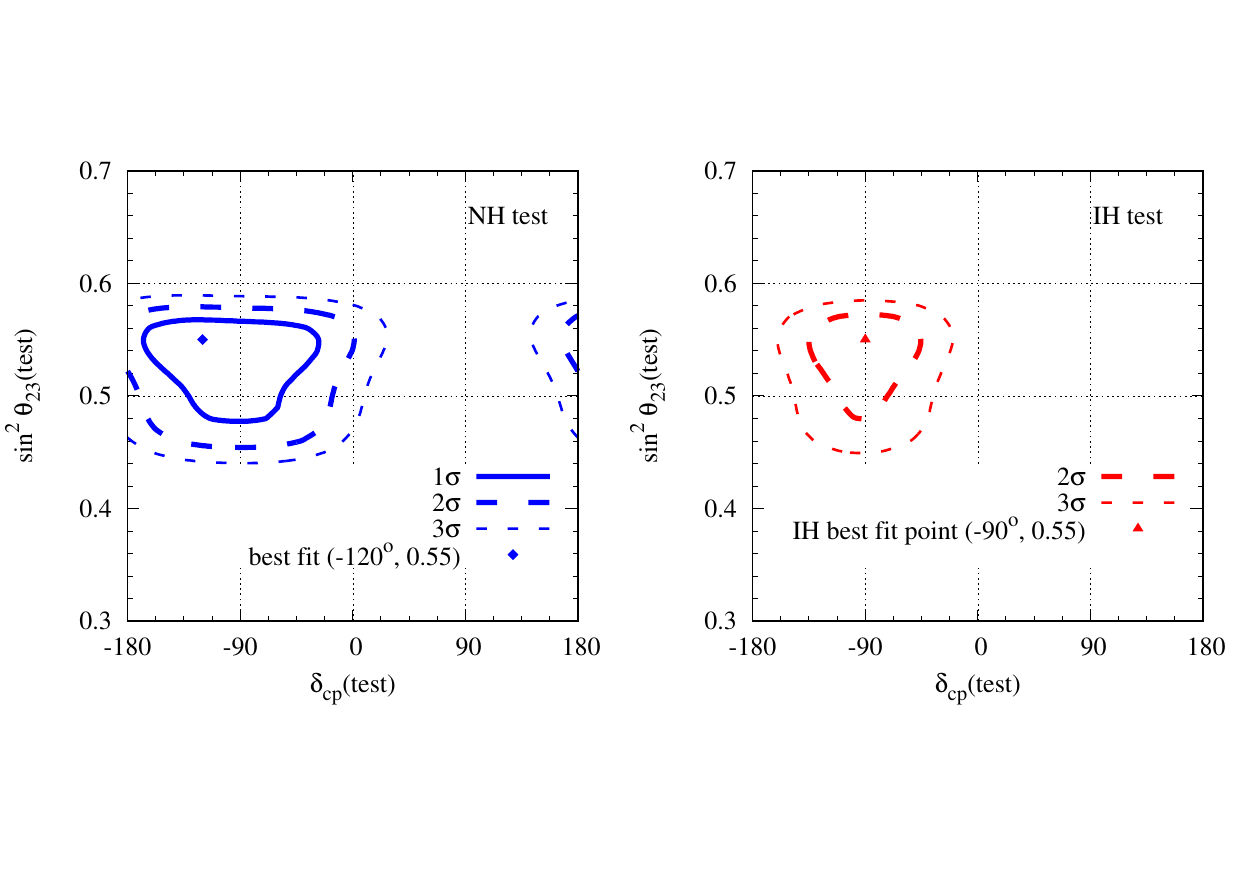}
\caption{\footnotesize{Expected allowed regions in 
$\dcp-\sin^2\theta_{23}$ plane from the combined fit of the neutrino and 
anti-neutrino data of T2K and NO$\nu$A, as of July 2019. In the left panel, the 
hierarchy is assumed to be NH and in the right panel, the hierarchy is assumed  
to be IH. The $\chi^2$ for NH best-fit point is $209$ and that for IH best-fit point is $211.5$, for $182$ 
data points.   
}}
\label{newnova19}
\end{figure}

Recently \nova collaboration published their results with increased anti-neutrino run~\cite{Acero:2019ksn}. The data 
in this analysis is based on $8.85\times 10^{20}$ POT in neutrino mode (which is the same for the previous analysis 
also) and $12.33\times 10^{20}$ POT in anti-neutrino mode (which is double that of the previous analysis). They have 
observed $27$ $\bar\nu_e$ appearance events. The results of 
this analysis give the best-fit point as (NH, $\sin^2\tz=0.56$, $\dcp=0$). The inverted hierarchy is disfavored with 
its best-fit point (IH, $\sin^2\tz=0.56$, $\dcp=-90^\circ$) being allowed only at $1.8~\sigma$. We see that the 
tension between the T2K data and the \nova data persists because the preferred values of $\dcp$ are widely different. 
We did a reanalysis of the 
data from T2K~\cite{Abe:2018wpn} and \nova~\cite{Acero:2019ksn}. The results are plotted in fig.~\ref{newnova19}. 
Comparing it with fig.~\ref{newsys10}, we note that the allowed regions have become more constrained though they are 
very similar to the previously allowed regions. We also note that the best-fit value of $\dcp$ changed from 
$-130^\circ$ to $-120^\circ$ and IH is allowed only at $2~\sigma$. Thus we see that the additional anti-neutrino data 
of \nova leads only to small changes in the combined fit.

\section{Conclusion}
The two long baseline accelerator neutrino experiments, T2K and NO$\nu$A, have taken significant amount of data both 
in the neutrino channel as well as in the anti-neutrino channel. The disappearance data of T2K prefers $\sin^2\tz$ 
close to $0.5$ whereas that of \nova prefers $\sin^22\tz$ to be non-maximal. T2K has observed $89$ $\nu_e$ 
appearance events but the number of $\bar\nu_e$ appearance events is not statistically significant. \nova has observed 
$58$ $\nu_e$ and $18$ $\bar\nu_e$ appearance events and has established $\bar\nu_e$ appearance at $4~\sigma$.

To understand the constraints imposed by the appearance data on the three unknown parameters of neutrino 
oscillations, we define a reference point: no matter effects, $\sin^2\tz=0.5$ and $\dcp=0$. We consider the change 
induced in $\pme$ and $\pmebar$ by the inclusion of matter effect due to NH/IH, by the change of $\sin^2\tz$ to HO/LO 
and by the effect of CP-violation with $\dcp$ in 
LHP/UHP. Both matter effects and non-zero $\dcp$ induced opposite deviations in $\pme$ and in $\pmebar$. But the 
octant of $\tz$ 
changes both the probabilities the same way. The observed $\nu_e$ appearance events in T2K are about $50\%$ more than 
what is expected for the reference point. Such a large excess is possible only if the change in $\pme$ is positive 
due to the changes in all the three unknowns. That is if hierarchy is NH, $\tz$ is in HO and $\dcp\approx -90^\circ$.
The best-fit point of T2K finds the unknowns to be: hierarchy is NH, $\sin^2\tz=0.53$ and $\dcp=-107^\circ$. The 
value 
of $\sin^2\tz$ is a compromise value of the best-fit values of the disappearance and the appearance data.  
T2K appearance data requires $\dcp$ to be in the neighbourhood of $-90^\circ$ quite strongly. In 
the case of NO$\nu$A, the observed $\nu_e$ and $\bar\nu_e$ appearance events are in moderate excess relative to the 
reference point. Such an observation can be explained only if the changes induced by hierarchy and $\dcp$ in $\pme$ 
and $\pmebar$ nearly cancel each other and the increase is due to $\tz$ being in HO. This is why \nova obtains two 
nearly degenerate solutions: hierarchy is NH, $\tz$ in HO and $\dcp\approx 30^\circ$ and hierarchy is IH, $\tz$ in HO 
and $\dcp\approx -90^\circ$. The large excess of $\nu_e$ appearance events in T2K rules out both these points at 
$95\%$ C.L. On the other hand, the moderate excess of $\nu_e$ and $\bar\nu_e$ appearance events in \nova disfavours 
enhancement of $\pme$ due to both hierarchy and $\dcp$.

The analysis of \nova data picks $\sin^2\tz=0.58$ as the best-fit value~\cite{Jun-2018}.
In the combined analysis of the appearance and the disappearance data of the two experiments the best-fit value of 
$\sin^2\tz$ is pulled a little lower by the disappearance data of T2K. The best-fit value of $\dcp$ for the NH solution is in 
the LHP at $-130^\circ$. This value is the result of the large excess of $\nu_e$ appearance events seen by T2K which 
force $\dcp$ to take a large value in LHP. Values of $\dcp$ in UHP, for NH, are 
ruled 
out at $2~\sigma$, even though the best-fit point of \nova is in this region. This also is a result of the large 
excess of 
$\nu_e$ appearance events observed by T2K. Values of $\dcp$ in UHP predict the number of $\nu_e$ 
appearance events for T2K  to be close to or below that of the reference point. Such values are strongly disfavoured 
by T2K because the observed number of events is significantly larger. Even though this region is preferred by
\nova appearance data, the conflict between its predictions and T2K data is ruling it out at $2~\sigma$ in the 
combined fit. 

Even though T2K barely allows an IH solution at $2~\sigma$, the combined fit has a nearly degenerate IH solution 
which is the common IH solution of each experiment, with $\dcp = -90^\circ$  and $\sin^2\tz = 0.56$. 
If the hierarchy is IH and $\dcp$ is in UHP $\pme$ is 
doubly suppressed by matter effects and by $\dcp$. There is a corresponding double enhancement of $\pmebar$. Such a 
feature is not seen 
by either experiment hence this possibility is ruled out at $3~\sigma$.

We have redone our analysis where we have included the latest \nova data~\cite{Acero:2019ksn}. The combined analysis 
of T2K plus \nova data leads to slightly smaller allowed regions with slightly shifted best-fit parameters. The 
tension between the T2K data and the \nova data still persists because the former prefers $\dcp$ close to maximal CP 
violation whereas the latter prefers $\dcp=0$. Our best-fit points, for both NH and IH, are close to the 
corresponding best-fit points obtained by the Nu-fit collaboration~\cite{nu-fit}, in their global 
neutrino oscillation data analysis, which includes the latest long baseline accelerator neutrino data. 
Therefore, a simple analysis of the data of T2K and 
NO$\nu$A experiments leads to reliable information on the values of the unknown parameters.

\section*{Acknowledgements}
SP thanks S\~ao Paulo Research Foundation (FAPESP)
for the support through Funding Grants No. 2014/19164-6 and
No. 2017/02361-1. UR thanks Department of Physics, IIT Bombay for the use of facilities and Industrial 
Research and Consultancy Center (IRCC), IIT Bombay for 
financial support.


\end{document}